\documentstyle[eqsecnum,aps,preprint]{revtex}
\include{psfig}
\begin{document}

\draft

\title{
Local state space geometry  and  thermal relaxation in complex landscapes: 
the spin-glass case.  
} 
\author{
Paolo Sibani\\
 }
\address{
 Fysisk Institut, Odense Universitet\\
Campusvej 55,\ DK5230 Odense M \\
Denmark
} 

\date{\today}
\maketitle

\begin{abstract} 
 A simple geometrical characterization of  configuration space
  neighborhoods of local energy minima in spin glass landscapes  
 is found by exhaustive search. 
 Combined with previous   
 Monte Carlo  investigations
 of thermal domain growth, it    allows a  discussion of  the connection 
 between real  and configuration  space  descriptions
 of   low temperature relaxational dynamics. 
 We argue that the part of state-space corresponding to a 
 single  growing domain is adequately modeled  by a   
 hierarchically organized set of states and that  
 thermal  (meta)stability in  spin glasses
 is  related to the  nearly exponential local
  density of states present within each trap. 
\end{abstract}


%\narrowtext
 
\section{Introduction} 
 In this paper we consider the    relation 
   between configuration space
  geometry, excitation morphology and relaxation dynamics   
  of  spin glass systems.  Spin glasses have 
   a complex landscape fraught with local (free) energy minima,
   which,   at sufficiently low temperatures, prevent the system 
  from reaching   a state of global thermal  equilibrium:
   slow relaxation   becomes 
 the dominating observable features of the system on relevant time
 scales,   and  thus  a main  point of theoretical interest.
 Despite the  absence of global equilibration, 
 spin glasses -- as well as   other complex systems displaying 
  broken ergodicity\cite{Palmer82}--
  relax through a series of quasiequilibria  involving   
  subsets of configuration  space  (traps or metastable regions). 
 Hence, during relevant  experimental time windows  measurable quantities  
 appear as  canonical averages  
 restricted to the configurations of the trap, and the latter can   be 
 considered  as a single point in the  free energy landscape of the system.
  Free energy barriers separating
 metastable regions can be   probed
 by  so called aging experiments\cite{Lundgren85,Alba86}:
 In   Zero Field Cooled (ZFC) experiments\cite{Mattsson94}, 
 the system is rapidly cooled in
 zero  field to a temperature $T_m$  and is then left 
 undisturbed for a time $t_w$.
 Thereafter, a   small  magnetic field is turned on, and the 
  growing  magnetic response of the system   is measured.
 In   Thermo Remanent Magnetization (TRM) experiments\cite{Lederman91}  the 
 system is  quickly cooled to $T_m$ and  allowed to  relax  in a 
 magnetic field for
 a time $t_w$. The field is then cut  and the decay of the 
 magnetization  is measured. In both cases the response curve  
  strongly depends on the time $t_w$  elapsed previous to the
  field change. Qualitatively, this can be understood
 by the idea of  partial equilibration just explained:  
after relaxing for a time $t_w$ the system
occupies a valley    in state space with a characteristic size
 related to $t_w$. Probing the system  on times scales less than $t_w$ reveals
 properties of the equilibrium fluctuations in this valley, while probing
 on longer time scales gives the non-equilibrium properties\cite{Sibani97}. 
 Numerically,  a coarse grained picture of configuration space
 relaxation can   be obtained directly for microscopic systems of 
  {\em modest} size 
  \cite{Sibani93,Sibani94b} by constructing
  the connectivity matrix  of a subset  of state space and
  solving the  master equation for the probability
  flow.
  For  each time, configurations mutually close to equilibrium
  are lumped  together, yielding a relaxation tree which describes
  the merging in time  of larger and larger subset of internally 
  equilibrated states.
  
 Theories  of spin glass dynamics are  either  
 based on  assumed  scaling properties of low energy excitations 
 in real space\cite{Bray87,Fisher88,Koper88},
  or on mesoscopic hierarchical models  of state space,  reviewed for instance
 in Ref.\cite{Vincent91}.
 Some of the approaches in  the latter category
   are inspired by    the equilibrium properties 
 of mean  field models\cite{Mezard87}, whose relevance for short-range spin
 glasses is now  hotly debated\cite{Newman96}. However,  hierarchical models    
 can  also simply be  viewed \cite{Palmer84}   
 as    paradigms of sequential, strongly constrained
 dynamics, without  reference to any equilibrium property. 
The models   generally feature a sequence of increasing barriers,
 delimiting larger and larger subsets of state space.  The simplest
 class of mathematically analyzable structures with this property are trees.  
 Master equations on 
  trees\cite{Ogielski85,Schreckenberg85,Grossmann85,Hoffmann85,Sibani86,Sibani91,Uhlig95}
 have  propagators which decay algebraically in time, and a  decay 
 exponent (or exponents\cite{Sibani91})  with a linear or s-shaped
  temperature dependence  ( depending on the particular  model ).
 When simple  `magnetic'   assumptions   are introduced,
  one can  reproduce the 
 AC susceptibility curves \cite{Sibani87}, the  aging behavior\cite{Sibani89,Hoffmann90,Joh96}, 
including changes of the apparent age due to a  temperature step\cite{Schultze91}
 and age  reset by temperature cycling\cite{Hoffmann97}.
 As the predictions of hierarchical models 
 are in good agreement with the experimental facts, a direct check  
 of their     basic  assumptions seems a worthwhile endeavor. We note that 
 as   low temperature excitations in spin-glasses are small non-interacting
 clusters in a frozen environment\cite{Dasgupta79,Sibani94,Andersson96},
  parallel and sequential modes  of  relaxation must  both    
 be  represented in the dynamics. This    
 leads to the question of exactly what parts of   state space, if any, 
  hierarchical models describe, and 
 to the allied, more general question, of  the relation between 
 state space phenomenologies and the widely used  domain growth approaches.

A numerical  state space analysis  is  attempted 
   here,  by  exhaustively  enumerating and statistically 
 analyzing  millions to hundreds of millions of configurations  
 close to deep energy minima.
  We do not try  to study the state-space
 dynamics directly,  since the  
 problem sizes considered prevent us from
 calculating connectivity matrices and numerical solving the master
 equation for thermalization. Exhaustive  search  is insensitive 
   to entropic barriers\cite{barriers} and gives  therefore
 incomplete information with regard  to dynamical properties. 
 However, by including  in the discussion  
  previous studies of  domain growth  and by using 
  simple heuristic arguments,   some conclusions can  be drawn about
  thermal (meta)stability.   
  
 The    extensive calculations  required in this investigation
 were done  on a Silicon Graphics
 Onyx parallel machine with 24 processors and 2 gigabytes of RAM at the
 University of Odense, Denmark, and on 
 a Silicon Graphics   32 CPU Origin2000 with up to 32 GB of RAM 
 kindly made available to this project by
 the  SGI Advanced Technology Centre in Cortaillod,
 Switzerland.
    
\section{Method and results}
Low temperature excitations in spin glasses are 
 small connected  domains of spins  which 
are reversed  relative to the their orientation in a 
very low  energy  configuration.
This  gives rise to a block structure in   state space,
where each block contains the dynamically available  states of  a single  domain.
Insight in the state-space geometry of one  block can  
 arguably be obtained  by considering   systems of  size comparable to a domain:
At low $T$  this   typically means   up to  $10^d$ spins, where $d$ is the 
dimension.  Even in such relatively small 
systems the  number of configurations quickly  becomes astronomical,
precluding the possibility of examining  {\em all } the configurations.  
Instead, one can use  the `lid method', which  exhaustively  visits
   relevant low-energy  parts of  the  landscape. \cite{Sibani93,Sibani94b}. 
A brief outline of the method is given here for completeness,
while a  more technical description, including details
of the algorithm  parallel implementation and 
performance statistics, is planned for a separate
 publication\cite{Sibani98}.  
 
 Consider a nearest neighbor walk 
 which starts at  a local energy minimum,  or reference state,    $\psi$,
 which, by definition,  has zero energy. All    states  with energy 
above an    energy `barrier',   $L$ are excluded from the
walk.  
We let  ${\cal P}_{\psi,L}$
be the set  of all states  which  can be  visited in such a walk. 
Following    our
 previous notation\cite{Sibani93} we refer to ${\cal P}_{\psi,L}$ as   
   a pocket of depth $L$ centered at  $\psi$.  
As,  in a thermal relaxation process,  the `typical' time needed
 to exit  ${\cal P}_{\psi ,L}$ (whether it be a first passage 
time or an average )   contains  the factor  $\exp (L/T)$, which
is  large at low temperatures,  ${\cal P}$ is a bona fide
candidate for a metastable region  operationally defined by an energy barrier.
 
 The reference state $\psi$ is constructed by an
  iterative process. Initially, we pick a random 
 state ${\cal \psi}_0$ to start the exhaustive search. As  soon as a new state
 with lower energy $\psi_1$ is found, it becomes the current reference
 states and  it is  used as a starting point  
 of a new search. At the $n'th$ stage of the process
 $\psi_n$ is the state of current   lowest energy among {\em all} 
  states examined.
 In this way the search algorithm `digs'    deeper and deeper
 into the landscape. The program stops when   a certain fraction of the
 spins in the system (usually one half) is  flipped without finding  
  any  state with energy lower than the current reference state. The latter is
   becomes  then, by definition,
   the `deep' minimum  $\psi$. 
 Our exhaustive approach considerably differs   from that of Vertechi and
 Virasoro\cite{Vertechi89}, who sampled  the distribution of energy 
 barriers  separating zero temperature 
 solutions of the TAP equation: in our case the barrier is  
 a parameter defining the pocket and  all states
 in the pocket are found deterministically, allowing the
 calculation of the local density of states.  
 
The spin glass problem deals with  a set  of $\cal N$ Ising spins
$\sigma_i ={\scriptstyle \stackrel{+}{-}} 1$ 
placed on a 2D or  3D cubic lattice with periodic
boundary conditions.\ 
 The energy of the $x$'th configuration,
$0< x \leq 2^{\cal N}$,     
 is defined by the well known Hamiltonian\cite{Edwards75} 
\begin{equation}
E(x)= - \frac{1}{2} \sum_{i,j} J_{ij} \sigma _i(x) \sigma _j(x),
\end{equation}
where  $J_{ij}= J_{ji}$  and where $J_{ij}\neq 0$
only if spins  $i$ and $j$ are   
adjacent on the grid.\ In this case,\ and
for $i<j$,\  we take the 
$J_{ij}$'s as  independent 
gaussian variables,\ with zero average  and
  variance $J^2 = 1$.\  
By definition neighbor configurations are  those
which differ  by  the   orientation of exactly one spin.

Data are  analyzed  with an eye  to the properties
of mesoscopic state-space  models. First and foremost we investigate,
for $E\leq L$,  
 the energy  distribution of the states     
or local density of states  ${\cal D}(E,L )$ within a pocket of depth $L$. The integral 
 of this quantity over its first argument   is the state-space volume of the pocket,
${\cal V}(L)$. In most cases, the reference  state $\psi$   has the property that  
 half of its spins can be overturned without finding any   
 state of lower energy. Let $L_{1/2}$ be the corresponding 
energy barrier. For the smallest
 examples considered this barrier was found (within a small tolerance)
 to  also allow  the system to reach the spin reversed state $-\psi$\cite{reversal}. 
  By chosing different initial condition for the search,  
 while keeping the $J_{ij}$ fixed, different deep reference minima can be found 
  which are unrelated by 
 global flips and which  all have energies per spin at or below
 estimates of the ground state energy  of spin-glasses\cite{Binder}.
  We also present data for more shallow pockets, i.e. 
 where the reference state is not particularly low-lying. 
 We furthermore calculate the barrier values for which the current 
 largest Hamming distance to 
 the reference state increases  its value, and  
 the average number of neighbors to which 
 a state is connected  as a function of the lid. This number would
 be constant and equal to the number of spins if all neighbors were 
 available, but as high lying states are discarded, the effective 
 connectivity is  much lower.
 
 All the investigations are performed for two and three dimensional lattices,
 which have similar patterns of low temperature relaxation\cite{Mattsson92}. 
 In both dimensions we analyze, for  a series of lattice sizes, 
  $25$  different realizations of the  $J_{ij}$'s. 
 As usual, disorder averages are performed. Furthermore  we 
 try  to convey an idea of the system to system and pocket to pocket
 variation. 
 In all cases we attempted to perform the exhaustive search up to  
 the barrier  value $L_{1/2}$. Due to memory limitations,
 this was unfortunately   not always possible
 for the {\em largest} among the  systems considered: 
 For the $10^2$ systems we succeeded in 24 cases out of 25, missing 
 by two spins in the unsuccessful case, while 
 for the $6^3$ systems we succeeded in 
   $20$ cases out of $25$. The remaining instances had to be stopped
 at different stages of the calculation.  
 The value of  $L_{1/2}$ and the size of the corresponding pockets
 greatly vary from sample to sample.   
 For ease of display, we  used the total volume of each pocket
 as a normalization factor for the    corresponding local density 
 of  states. This creates   vertical  shifts of the data 
 which are void  of physical significance.

Before considering any issue of interpretation, 
 it is interesting to note that all the data can be parametrized rather
   simply: the
 local density of states is almost exponential,
 with a  small systematic downward curvature (see Figs. 1-4).   
 The available volume in the pocket is also close to an  exponential
 function of the lid energy, if one smooths out  the jumps which
 occur  when a new `side pocket' appears as the barrier increases
 (see Figs. 7 and 8). Finally, the largest obtainable Hamming distance
 to the reference state grows exponentially with the barrier
 defining the pocket(see Figs. 5-6).
 All these  properties are common to  deep and less deep pockets.
  
The exponential form of the local density of states
is shown directly in Figs.  1 and  2 for  
$10^2$ and $5^3$ systems respectively. 
In these plots  the raw data are indicated
by different symbols, one for each realization
of the couplings, while the  lines are calculated  by  fitting
 $\log( {\cal D}(E,L_{max}) ) $ to a parabola
in $E$.  The coefficient of the second
order term is typically 20 times smaller than the coefficient
of the linear term in 2D,  and 50 times smaller in 3D.
Therefore, the reciprocal of the linear  coefficient   basically describes  
the energy  scale of the exponential density of state at low energies. 
These   coefficients are  plotted versus the  system size in Figs. 3 and 4
  for a number of 2D and 3D systems respectively.
 As discussed later, 
they  provide an estimate  of the   
kinetic transition temperature $T_k$ at which  the system looses
its thermal (meta)stability\cite{Andersson96}. 
 
 Figs. 5 and 6   describe 2D and 3D systems respectively. 
 The ordinate is the Hamming distance to the reference state,
 (scaled for convenience to the unit interval), while the abscissa is
 the average energy barrier
  which must  be overcome to achieve that Hamming distance.
  Each data set   represents a different system size, and in each
  case the average is performed over 25 realization of the disorder.
 
In 3D we additionally  performed two different  types of investigations:
 1) for one particular 
realization of the $5^3$ systems,  we varied the initial conditions of
 the search leading to the reference state and 
2) for one particular realization of
a $6^3$ system we studied the local density of states for `shallow'
 minima. 
 As a  few percent of the states
are energy minima\cite{Sibani93,Sibani94b}, 
it is not surprising that  different  {\em quenches}   lead 
 to different final states.
It is however  interesting that  careful optimization 
 similarly leads to   different reference states, which are
 unrelated by symmetry and separated by  barriers
 $\geq L_{1/2}$.  This is shown in   Fig. 7,
where  the volume $\cal V$  of three different deep pockets (for the 
same $J_{ij}$) is plotted vs. $L$. To 
avoid partly overlapping data points
we multiplied the data by different factors, as indicated in the 
figure itself.

In the second investigation we consider a sequence of
five shallow pockets:   The first data set   is 
generated by  starting the search  in    a  high lying minimum,
and stopping it  as soon as a lower minimum is found. This
newly found minimum  then serves as a starting point for the second stage
of the search,  leading to the states in 
 pocket $\# 2$, and so on. 
 Figure 8  shows the local density of state found in each of those
 pockets. The local density
 of states as well 
  as the  volume   (not shown)   are still  
 close to exponential functions of their arguments. Indeed, this behavior
  appears  in all the pockets we examined,
 regardless of the energy of the reference state.  
  
\section{Discussion} 
  
Most  theoretical descriptions of spin glasses either use     
insights and numerical observations  on the morphology of excitations
at low temperature\cite{Bray87,Fisher88,Koper88,Dasgupta79}  or 
build on   ideas  about  the structure of state-space
\cite{Lederman91,Sibani87,Sibani89,Hoffmann90,Schultze91,Hoffmann97,Campbell86}.  
The first method  is rooted  in
 a renormalization group approach developed by   
Mc Millan\cite{McMillan84} and Bray and Moore\cite{Bray85}.
It includes  droplet\cite{Bray87,Fisher88} and domain \cite{Koper88} theories,
which   emphasize  correlation  functions and their temperature and
field dependence. 
The early work of Dasgupta et al.\cite{Dasgupta79} is mainly numerical, but 
already  contains interesting theoretical observations  on the 
excitation morphology.
Other approaches are   either inspired\cite{Lederman91} by  
the equilibrium solutions of the mean field model \cite{Parisi83} 
or use a   purely dynamically motivated  lumping procedure to produce 
hierarchical models of state-space, within the general approach of 
 broken  ergodicity\cite{Sibani97,Sibani91,Uhlig95,Sibani87,Sibani89,Hoffmann90,Schultze91,Hoffmann97}.
 Campbell's approach\cite{Campbell86}  describes  the spin-glass transition
as  a  percolation transition in state space.
The hierarchical scenario described  by  
by Lederman et al.\cite{Lederman91}  and  Vincent et al.\cite{Vincent95}
 shares  with the  models of 
refs.\cite{Sibani87,Sibani89,Hoffmann90,Schultze91,Hoffmann97},
the idea  that, as the temperature is lowered, more and more
details of state space emerge. The difference is 1) in the interpretation
of the `vertical' direction in the tree, which is a magnetic overlap 
in one approach and an  energy in the other, and 2) in the  importance  attached to the
local density of states.
In  the approach  of Joh et al.\cite{Joh96},   
an ultrametric tree    describes   a manifold  of states of 
constant magnetization, and the   response behavior
to a field step $0 \rightarrow H$  is associated to a transfer of probability from the
$M=0$ manifold to a manifold  of states having the field   cooled
magnetization $M_{FC}$.  
A phenomenological treatment of relaxation in spin glasses introduced
by Bouchaud\cite{Bouchaud92} builds on the  idea that the 
 distribution of barrier heights within a trap is exponential. 
 Even though the approach    differs from that of  models
where   the barrier    {\em parametrizes} 
 state space, an underlying  similarity is that,  in any  tree structure with a 
minimum of regularity, the distribution of the barriers 
delimiting subtrees is indeed typically  exponential.
 
Domain theories and hierarchical models
 have mainly  been regarded as competing approaches\cite{Vincent95,Lefloch92,Andersson94},
and the discussion over their relative merits has been intertwined with a 
a continuing  debate on the nature of equilibrium states of
 spin-glasses\cite{Newman96}. Even though the latter question has  great theoretical interest,
 low temperature relaxation never  probes    equilibrium properties, as for example 
indicated by the fact that similar  physical behaviors 
(e.g power law relaxation and aging) characterizes system of
 different dimensionality.       
We would argue    that  much of the low temperature phenomenology can 
be understood by looking at the physics on small length   scales, 
that  the  dichotomy between real space
 and configuration
space descriptions is moot   and 
 that (some)   hierarchical approaches are  justified regardless of 
the relevance of mean field behavior  for short range spin-glasses.

Under the assumption that the set of spin configurations available
to a domain is similar to the configuration space of a    
  small systems,  the growth of the average domain size $V_{rs}(t)$ 
 should match the  growth
 of the configuration  space volume $V_{cs}(t)$, which can be
 visited by a Metropolis algorithm in $t$ time steps.   
  At low temperatures, the former quantity grows 
    algebraically in time 
and can be written  as $V_{rs}(t) \propto \exp (b(t,T))$, where 
 $b$ is   the  barrier $b(t,T) \approx T\ln(t)$ which thermal
 fluctuations typically overcome in time  $t$\cite{Andersson96}. 
We   have unpublished evidence that 
$V_{cs}$  grows algebraically as well. The same  conclusion
can   be reached from the present evidence, if one  makes the
( reasonable ) additional   assumption  
that, starting from 
zero energy, states of energy $b$ are visited  on
a time scale $\alpha(T) \ln t$. (An  Arrhenius behavior 
corresponds to $\alpha(T) \propto T$, and we  
expect   $\alpha(T)$ to be close to a linearly growing function of $T$).
Since, as our present data show,  the volume of the configuration space
`below' $b$ increases exponentially 
with  $b$,  we  recover  $V_{cs} (t) \approx t^{C \alpha(T)}$,
where $C$ is some  constant.
Associating the configuration  space volume with  the
spins in a cluster thus implies  that
$V_{cs}  \propto V_{rs}^\beta$ for some constant $\beta$,  
which again implies a strongly  
constrained and predominatly sequential   dynamics within a cluster.  
We recall  that   several   very low energy configurations unrelated
by flip symmetry exist, as clearly shown by Fig. 7. 
  This is  somewhat  reminescent of  mean-field behavior, and consistent with 
 the results of  numerical investigations in small systems\cite{Reger90} showing  
a non-trivial    Parisi overlap function.  
It  could  also  be
related to the  chaotic  
 sensitivity  of  thermal correlations\cite{Bray87} in spin glasses: 
Since the barriers among    multiple deep 
minima are  high,  thermal fluctuations  are confined 
to the surroundings of  one specific state.
  However,  field or temperature changes  which destroy 
 local equilibrium and change the barrier structure can   
 induce   transitions from    one deep  pocket 
 to  another, leading,  in real   space,    to a 
partial obliteration  of the pre-existing thermalized
  domain structure and thus to thermal chaos.   

Let us  now  consider  how our 
present numerical evidence specifically relates to  
hierarchical models. It must be born in mind that these models 
 are  intended as a  
coarse grained  description  of state space:
some of their features 
i.e.  the lack of translational invariance,  the form of the 
local density of states, and the relation
between the  barrier and  the Hamming distance to the
 starting configuration, are of central importance
 to the relaxation behavior. By way of contrast,
the lack of  loops characterizing tree graphs  
only approximates the sparsity of the connection
matrix  and  cannot be an exact property  
of  the state space of  microscopic models. 
A   tree  obtained by successive branchings 
 from a `top'  node 
can   be viewed   as 1) a hierarchy of energy or free
energy barriers, with all 
the states lying at the  lowest level \cite{Ogielski85,Schreckenberg85},
 {\em or}  
2)  all nodes in the tree may represent  lumped 
physical states\cite{Grossmann85,Hoffmann85,Sibani86,Sibani91}. 
In the first case (and not in the second)     
 the system has an  ultrametric distance: 
for two arbitrary   bottom nodes  this  distance   is 
just  the number of levels  up to the top of the smallest subtree
containing both nodes. In mean-field models one identifies  
 the index of the hierarchy  with a   magnetic overlap, 
  while in the approach  developed 
in Refs.\cite{Sibani87,Sibani89,Hoffmann90,Schultze91,Hoffmann97},
the index of the hierarchy   is  an energy  
barrier\cite{entropic_barr_in_mod}, and  the statistical weight
  attached   the lumped nodes
increases exponentially with this barrier. This last feature agrees  with  the 
fact that  both the  local density
of states (Figs.~1 and 2)  and the state-space volume of pocket (Fig.~7) grow 
exponentially with their respective arguments. 
The magnetic properties of the   models hinge 
on  the  disorder averaged overlap with the reference state, which  
was   assumed  for simplicity to increase linearly along 
the `bottom' nodes  of the tree\cite{Sibani89}. The assumption 
 implies  that the
largest achievable Hamming distance within a 
pocket  should grow exponentially with the  confining   barrier.
As shown in Figs.~5 and 6, such an exponential relation is confirmed by the present
investigation. Let us finally  consider
the issue of thermal metastability, which, 
 as first noticed  in refs.\cite{Grossmann85,Hoffmann85},
 characterizes  models with an an exponential density of states 
 $\propto \exp(E/T_k)$.
At the `glass transition temperature' $T=T_k$ the equilibrium probability becomes strongly
 biased towards high lying, rather than 
low lying configurations,  basically 
due to  a change of sign in the   exponent  of 
$P(E) \propto \exp (-E/T + E/T_k) $.
If, as we here suggest, these models
describe the physics of a  trap, this trap  
must cease to be a thermal attractor 
 when the temperature exceeds $T_k$.
Clear  concurring evidence    is the dramatic change
in  domain growth\cite{Andersson96}, 
 from slow to fast, taking place at  a well  defined temperature
 - which  in 3D is close to $0.8 J$ ( and close to 
  the actual critical temperature ), while   
  in  2D it is close to  $0.6J$.  The change 
is  not simply related to the existence or absence   of a true equilibrium transition,
since    
it  appears already at short times and independently of dimensionality.  
Figures~3 and 4  display, as a function of 
the linear size of the lattice,  the `kinetic transition temperatures'
naively calculated as the reciprocal  of the scale of the
exponentially growing density of states,  for  a number  of realizations
of spin-glass systems in two and three spatial dimensions.  
If the confining barriers of the system were purely energetic, the 
density of states would contain all relevant 
information. The $T_k$'s would then be independent of the
system  size  and   coincide with  the kinetic transition
temperatures.
In reality, while the range of the estimated $T_k$ includes
the correct values,   the data have a clear downward trend,  leading 
to the expected conclusion that entropic confinement becomes important
 as the linear size of the system grows beyond  a certain limit, 
 in this case $\approx 4-5$.
What   likely   happens in a large system is that, when   
$T_c$ is approached from below, {\em small}  clusters melt  and start 
to interact strongly with one another other via the large
 amount of loose spins generated  in the process. 
 According to Fig.~4 this  would happen, for $4^3$ systems,
at $ T\approx 0.8$.
 
In conclusion, we have argued  that 1) the state  space configuration 
 corresponding to
a small isolated domain can be adequately described by  
a hierarchically organized set of states,    and 2) that  
the loss of metastability
of these  domains   is due to 
their exponentially growing density of states
which   plays an important,  albeit  not exclusive,  role
for the thermal  metastability of  
 system as a whole.\cite{Ising_ferro}.
Similar results appear to hold for other complex systems with multiple
local energy minima, as shown by the previous analysis of the  TSP problem\cite{Sibani93}   
and by ongoing investigations of  covalent network models of glasses\cite{Schon97} 

\noindent {  \bf  Acknowledgments:}  I am indebted to Richard Frost 
  of the San Diego Supercomputing Center   for  good advice on the
  principles of  parallel
computing and to   Ruud Van der Pas of the European High Performance Computing Team, SGI,
 for his feed-back during the development phases of the code, for his
 prolonged  assistance in performance tuning   and for actually 
 running the most memory intensive calculations on    hardware 
 which was kindly put at the projects disposal by   Silicon Graphics
Advanced Technology Center in Cortaillod, Switzerland.
 It is a pleasure  to thank Eric Vincent, Richard Palmer,
and  Christian Sch\"{o}n for comments on this work.
I would  also like to thank Peter Salamon  and the department of Mathematical Sciences
 at San Diego State University for   the  nice hospitality 
during my sabbatical leave of absence, the
  Santa Fe Institute of Complex Studies where part of this work was written,
  and the Danish National Research Council 
  ( Statens Naturvidenskabelige Forskningsr\aa d )
 for  generous financial   support.

\newpage
\begin{figure}
\centerline{\psfig{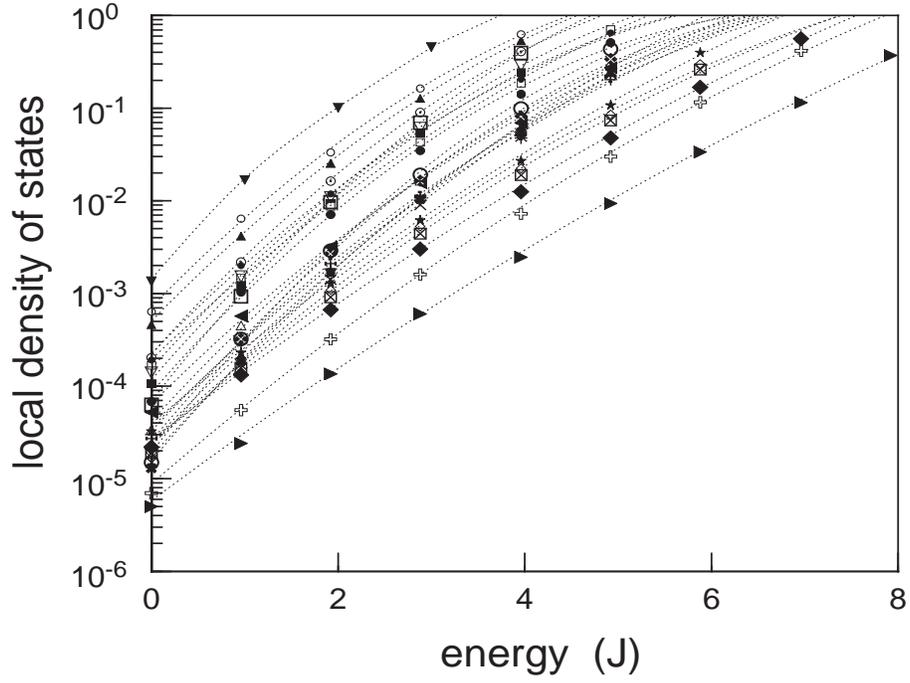}}
\vspace{1cm} 
\caption{  
The local density of states within a  pocket
centered at a deep energy minimum is shown as a function of the
energy in a semilogarithmic plot, for 25 different realizations
of the couplings in a $10 \times 10$ system.  
The dotted  lines are fits of 
 $\log \cal D$ to a parabola.  
 As we have   normalized  the integral of the density of states
  to one, the  differences in magnitude among the data sets
  directly  reflect  the   size of the corresponding pockets.     
}
\end{figure}
\newpage
\begin{figure}[t]
\centerline{\psfig{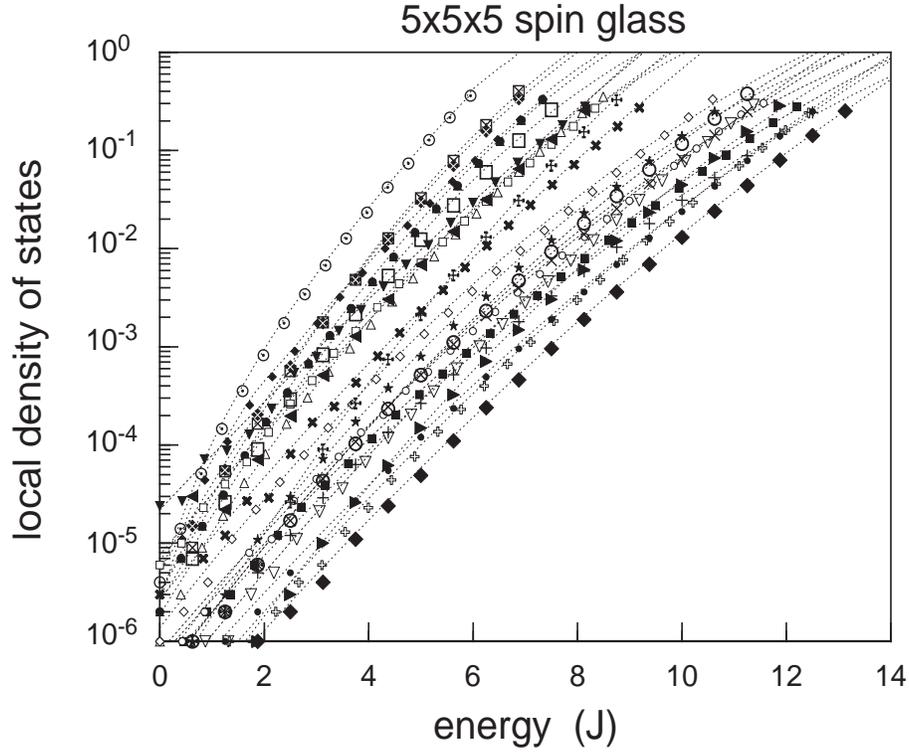}}
\vspace{1cm}
\caption{ 
The local density of states in a   pocket
centered at a deep energy minimum is shown as a function of the
energy in a semilogarithmic plot, for 25 different realizations
of the couplings in a $5 \times 5 \times 5$ system. 
 The dotted  lines describe fits of 
 $\log \cal D$ to a parabola. 
 As   the integral of the density of states is normalized 
  to one, the   differences in magnitude among the data sets reflect 
  the    size of  the pockets.     
}
\end{figure}
\newpage
\begin{figure}
\centerline{\psfig{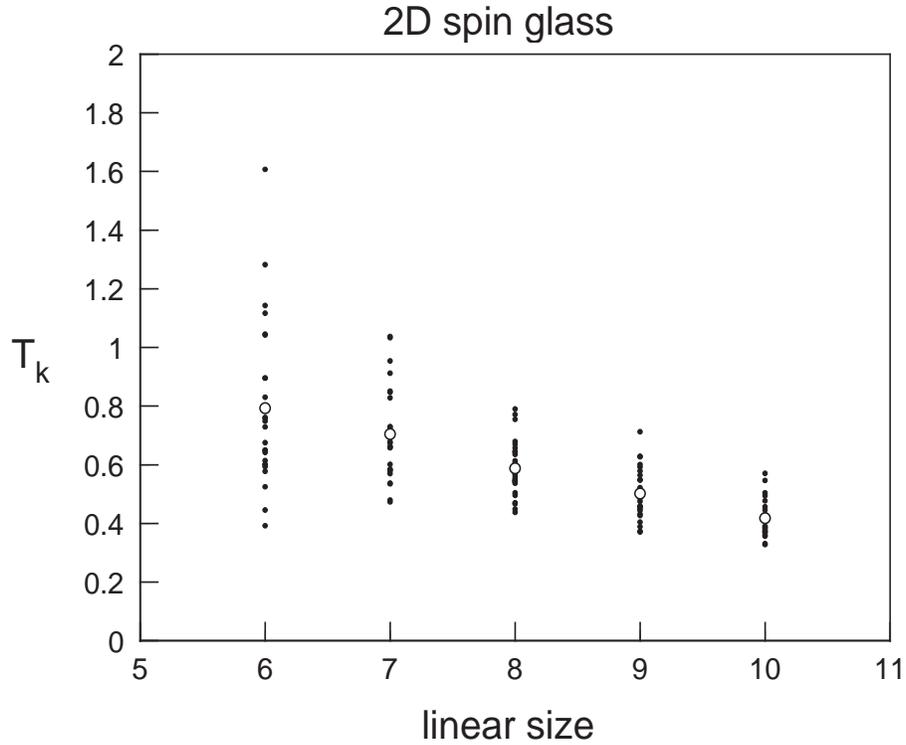}}
\vspace{1cm}
\caption{
Each     dot  marks  the estimated 
kinetic transition temperature $T_k$  of one realization of 
a 2D system of a given linear size.
 The average value of $T_k$ is indicated by  a circle.   
The  $T_k$'s are calculated by
 expanding   $\log \cal D$ to second order in  $E$
and taking  the reciprocal 
of the coefficient of  the (dominating) linear term. 
} 
\end{figure}

\newpage
\begin{figure}[t]
\vspace{1cm}
\centerline{\psfig{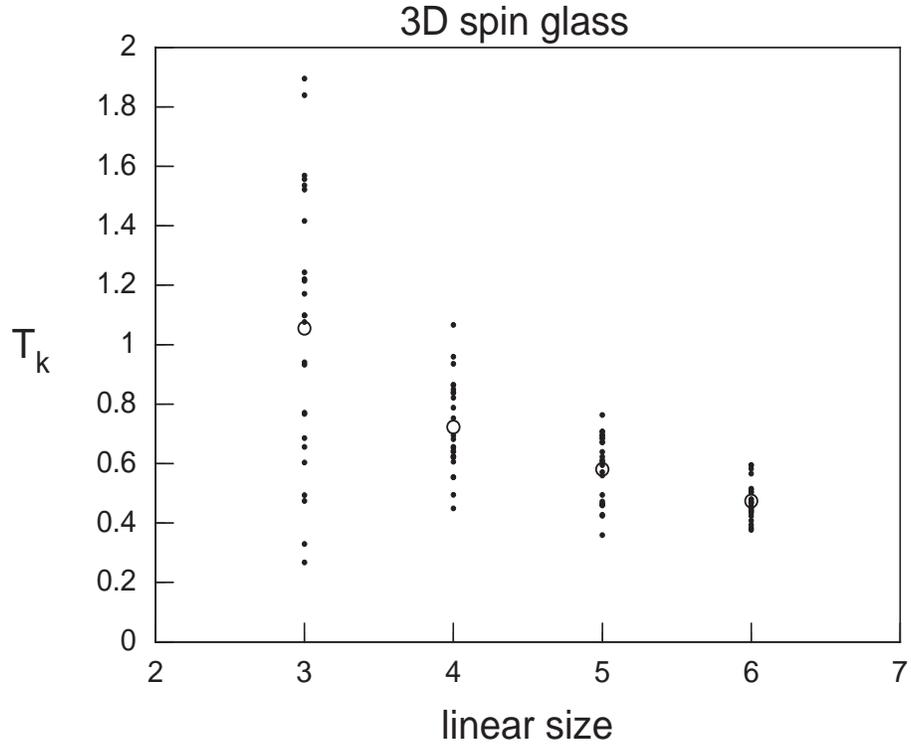}}
\vspace{1cm}
\caption{
Each  dot  marks  the estimated 
kinetic transition temperature $T_k$  of one realization of 
a 3D system of a given linear size.
The average value of $T_k$ is indicated by  a circle.
The $T_k$'s are calculated by   expanding
  $\log \cal D$ to second order in  $E$
and taking the reciprocal 
of the coefficient of  the (dominating) linear term.  
}
\end{figure}
\newpage
\begin{figure}[t]
\vspace{1cm}
\centerline{\psfig{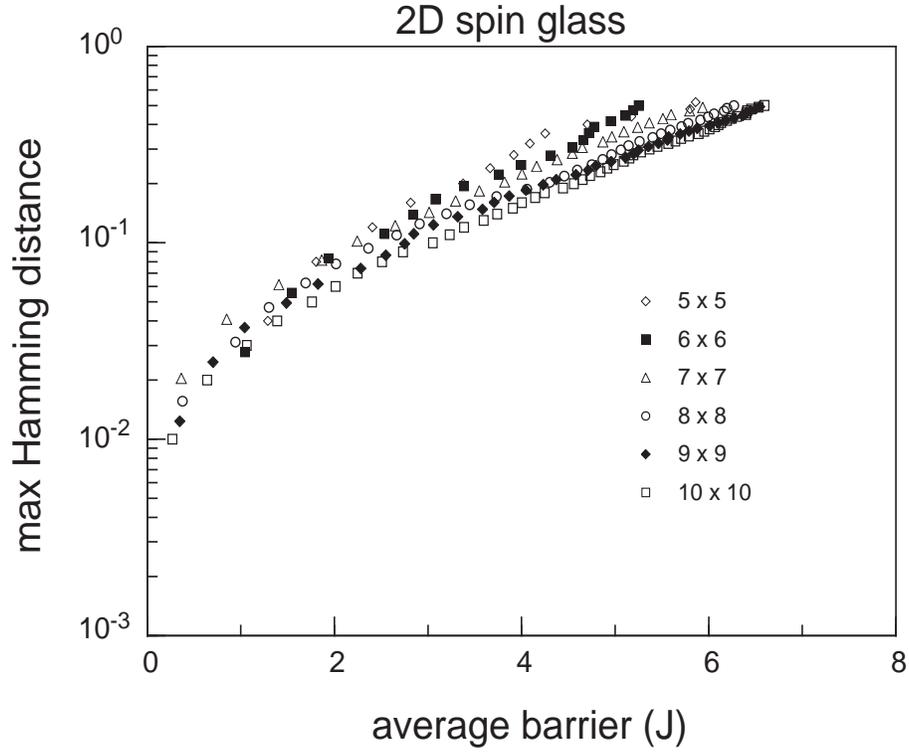}}
\vspace{1cm}
\caption{ 
The ordinate is the largest
 Hamming distance to   a reference state of very low energy 
which can be achieved  without overcoming the  energy
barrier given by   the abscissa. The data  are 
averaged over $25$ different disorder realizations of
a 2D system. 
Note that after a short transient, the dependence
is simply exponential.  
} 
\end{figure}
\newpage

\begin{figure}[t]
\centerline{\psfig{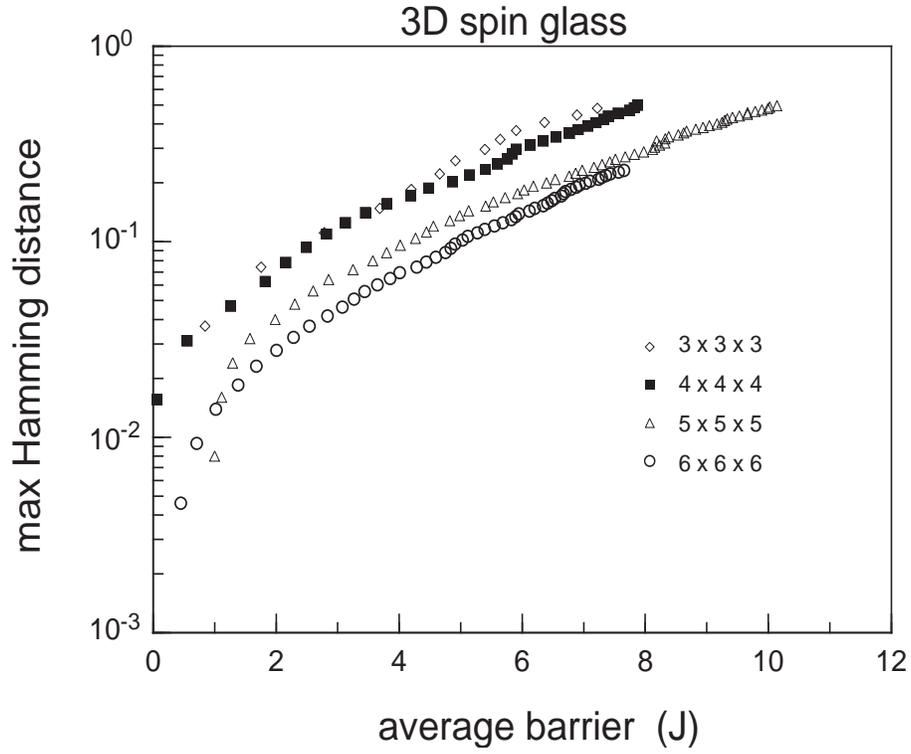}}
\vspace{1cm}
\caption{
The ordinate is the largest
 Hamming distance to   a reference state of very low energy 
which can be achieved  without overcoming the  energy
barrier given by   the abscissa. The data  are 
averaged over $25$ realizations of a
3D system. 
}
\end{figure}
\newpage
\begin{figure}[t]
\centerline{\psfig{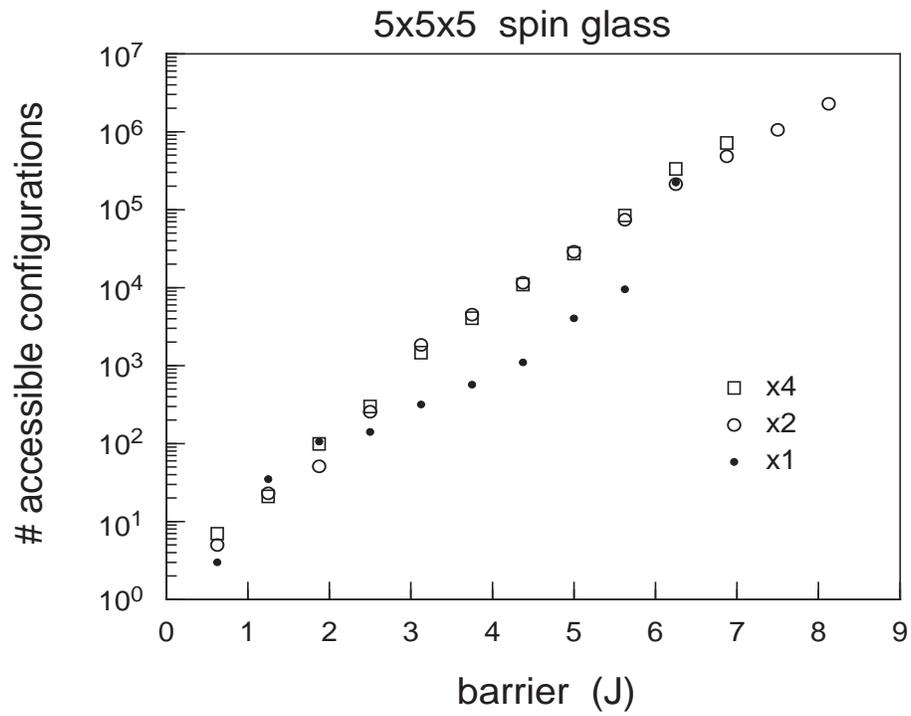}}
\vspace{1cm}
\caption{
The plot shows  the  characteristic exponential dependence of the 
accessible number of states  on the energy barrier for three
  disjoint pockets belonging to the same realization of the
  couplings in a $5^3$ system.    For convenience, the data have been 
multiplied by different factors as indicated in the figure itself.
} 
\end{figure}
\newpage 
\begin{figure}[t]
\centerline{\psfig{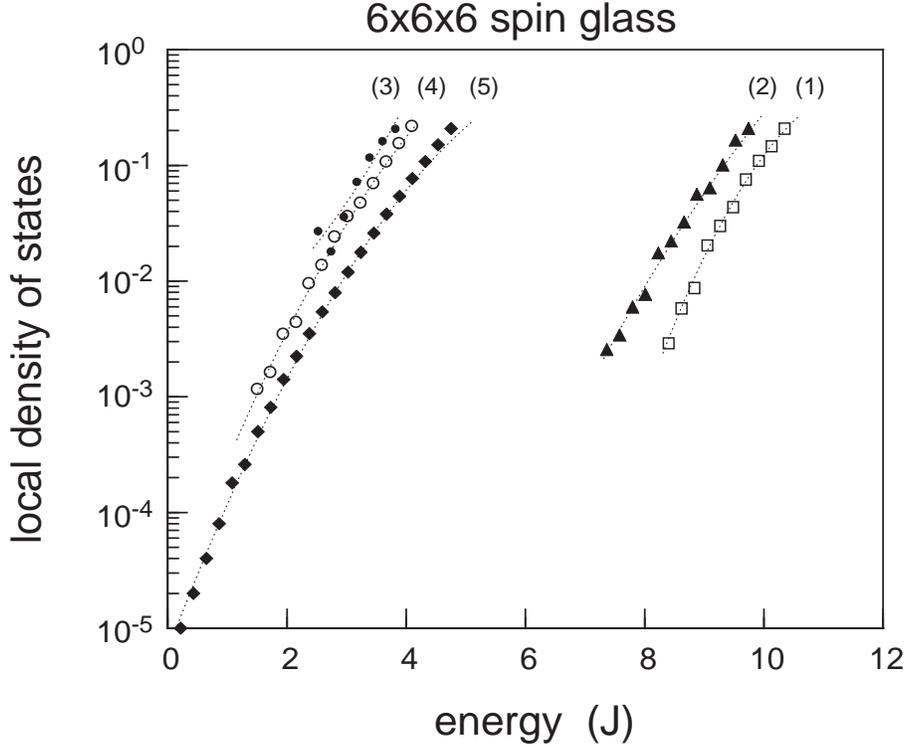}}
\vspace{1cm}
\caption{
The  plot describes the local density of states in a series
 of `shallow' pockets  belonging  to a single $6^3$ system.
  The rightmost  data set is obtained by starting the 
 exhaustive search in a  rather poor local energy minimum 
 (energy   $= -1.605716 J/$spin ) and stopping it  as soon as
 a lower energy minimum is found. This  new minimum 
   (energy  $=  -1.610523 J/$spin ) 
 is the starting point for  the  search producing  the 
 second set of data, which again stops when a better minimum is
 found. All shown data sets are obtained in a similar fashion. 
 The values of the energy per spin  of the reference states  
  -- besides the  first two
 just mentioned  -- are  
 $-1.632942,  -1.637671$ and  $-1.643615J/$spin. 
  For convenience  
 the  energy of the very lowest
 energy configuration is used   as the  origin of the abscissa. 
The densities are normalized  by  the total number of states
  in the corresponding  pocket. 
}
\end{figure}


\begin{references}
\bibitem{barriers}
To  appreciate the significance of the entropic barriers, consider  a 
   simple ferromagnetic Ising model, which 
   only has  two equivalent  energy minima.
  The energy barrier between them is  associated  to  a domain wall,
  and  scales as the linear size of the system. As  the  mean energy is extensive, 
  it is an entropic barrier, i.e. the very small likelihood of  focusing this energy in the right place, 
   which accounts  for the lack of ergodicity in the limit of large system size.
\bibitem{reversal}
In general, we expect that 
 the reversal barrier $L_{rev}$ should be quite close to  $L_{1/2}$. This can be seen 
 as follows: by symmetry  the lowest barrier needed  to flip half of the
 spins will be the same, irrespective of whether  one starts from one deep minimum or its
 mirror image. These  two  flipping procedures  
 yield   excited configurations  which are each others mirror
 images. If   they could be connected to each other with no further 
 energy expenditure,   then $L_{rev} = L_{1/2}$ would follow. Unfortunately,
 checking this property numerically seems out of the question for all but 
 the smallest systems. However, since both  excited 
 configurations have an energy which is  rather 
 high   compared to that of  single spin flips,
 it seems reasonable that a path between the two
 states  can be constructed by  matching pairs of consecutive
 flips with energy contributions of opposite 
 sign and nearly equal magnitude.
\bibitem{entropic_barr_in_mod}
In the simplest tree  models
the  barrier  confining   a subtree  is purely energetic.
However,  entropic 
barriers between  any two given states, which shape
the diffusion profile and the relaxation
exponents,  are also present. This is  due to the fact that  upon overcoming 
 the  energy barrier between the  two states,
 an  exponentially large number of other  low
 energy states also becomes available. 
We also note that the kinetic transition in the models is
itself  a manifestation of an entropic effect:  
 the overwhelming majority of high energy states
wins over the energetic part of the Boltzmann factor.
\bibitem{Ising_ferro}
As a check, the lid method was applied to the neighborhood 
of the ground state of an Ising 
ferromagnetic model with simple square or cubic lattice. 
 We analyzed a $7\times 7$, an  $8\times 8$ and a $4\times 4 \times 4$
  system and 
 counted $2 \times 10^5$, $8 \times 10^5$ and $9 \times 10^6$ configurations
 respectively.
In all these cases    the local density of states is  almost 
indistinguishable from a simple  exponential through the 
range of (low) energies
considered. The   $T_k$'s roughly estimated  as the energy scale
in this exponential are $2.66$, $2.70$ and $4.66$ $J$ respectively.
This must be compared with the true critical temperatures, which are
$2.27 J$ in two dimensions\cite{Ma} and $T_c \approx 4.51 J$\cite{Domb74} 
in three dimensions.
\end{references}
\end{document}